\documentstyle[aps,epsf]{revtex}

\newcommand \beq {\begin{equation}}
\newcommand \eeq {\end{equation}}
\newcommand \boldsigma {\mbox{\boldmath $\sigma$}}
\newcommand \boldxi {\mbox{\boldmath $\xi$}}
\newcommand \rra {\rangle \rangle}
\newcommand \lla {\langle \langle}
\newcommand \bay {\begin{array}}
\newcommand \eay {\end{array}}

\begin{document}

\title{Multifractality and percolation in the coupling space of perceptrons}
\author{ M. Weigt  \thanks{martin.weigt@physik.uni-magdeburg.de} and
        A. Engel \thanks{andreas.engel@physik.uni-magdeburg.de} \\%[-.2cm]
      {\small{\it Institut f\"ur Theoretische Physik,
      Otto-von-Guericke-Universit\"at Magdeburg}}\\%[-.4cm]
      {\small{\it PSF 4120,
      39016 Magdeburg, Germany}}}
 
\date{\small June, 4, 1996}
\maketitle
 
\begin{abstract}
The coupling space of perceptrons with continuous as well as with binary
weights gets partitioned into a disordered multifractal 
by a set of $p=\gamma N$ random input patterns. The multifractal spectrum 
$f(\alpha)$  can be calculated
analytically using the replica formalism. The storage capacity and the
generalization behaviour of the perceptron are shown to be related to
properties of $f(\alpha)$ which are correctly described within the replica
symmetric ansatz.  
Replica symmetry breaking is interpreted geometrically as a transition
from percolating to non-percolating cells. The existence of empty cells
gives rise to singularities in the multifractal spectrum. 
The analytical results for binary couplings
are corroborated by numerical studies.\\[0.5cm]
\end{abstract}

\section{Introduction}
Simple networks of formal neurons with emergent properties for information
processing have been discussed within the framework of statistical mechanics
for meanwhile more than 10 years. In particular the simplest case of a
feed-forward neural network, the single-layer perceptron, has been analyzed
from various points of view and with respect to rather different properties
in numerous papers. This is mainly due to the fact that the storage as
well as the generalization abilities of this network can be concisely 
described
using the phase space formalism introduced by Elizabeth Gardner
\cite{gardner}. Part of these investigations are summarized in
recent reviews \cite{OpKi,WRB}. 

On the background of an ever growing body of investigations aiming at more
and more special aspects of this system it seems appropriate to look for an
unifying framework that allows to characterize the various properties in a
coherent fashion. In the present paper we show that the geometrical
structure of the coupling space of the perceptron shattered by a random 
set of inputs offers such a possibility. In fact the statistical properties of
the partition of the coupling space into cells corresponding to different
output sequences can be quantitatively characterized using methods from the
theory of multifractals. With the help of the replica trick the multifractal
spectrum can be calculated explicitly. Many of the relevant properties of
the perceptron such as the storage capacity, the typical volume of the version
space and the generalization ability are
closely related to special properties of this multifractal spectrum. As a
result the relations between different investigations become more 
transparent. 

The idea to characterize the perceptron by the distribution of cells in
coupling space induced
by the inputs is rather old. It is already the basis of the classical
determination of the storage capacity by Cover \cite{cover} and is frequently
used in studies of information processing in mathematical statistics
and computer science (see, e.g., \cite{HKS}). Its qualitative
appeal within the framework of statistical mechanics was emphasized 
by Derrida et al. \cite{DGP} who, however, seem not to have realized that
the relevant quantities could in fact be calculated. This became clear only
after the work of Monasson and O'Kane \cite{MoKa} characterizing the 
distribution of internal representations in the reversed wedge perceptron.
Meanwhile these investigation were extended to the case of
multi-layer networks and have produced several new results \cite{MoZe,Cocco}. 
But also for the simple perceptron this formalism offers the possibility of a
systematic and coherent description clarifying several delicate points of
former investigations. In the present paper we present a detailed
analysis of the perceptron from this point of view. Some of the results
were already published in \cite{EW}.

The paper is organized as follows. In section II we present the general 
formalism
of multifractals in its application to neural networks. Section III  
contains the analysis of the spherical perceptron, 
in section IV the Ising perceptron
is discussed. A summary is given in the last section.

\section{General formalism}
In this paper we are going to analyse the coupling space of simple
perceptrons. These are defined by the relation 
\beq
\sigma = \mbox{sgn} ( {\bf J}\boldxi ) = \mbox{sgn} ( \sum_i J_i\xi_i ).
\label{perc}
\eeq
between $N$ input bits $\xi_i = \pm 1, \, i = 1,\ldots ,N$, and a single
output $\sigma = \pm 1$. We are interested in the thermodynamic 
limit $N\to\infty$. The coupling vector ${\bf J}\in I\!\!R^N$ is 
model-dependent: For the {\it spherical perceptron} the only condition is the 
normalization of this vector to $\sqrt{N}$, in the case of the {\it Ising
perceptron} it has binary components $J_i = \pm 1$.

We choose $p = \gamma N$ random independent and identically distributed 
input patterns
$\boldxi^\mu \in I\!\!R^N,\, \mu = 1,\ldots ,p$. The hyper--plane orthogonal
to each of these patterns cuts the coupling space into two parts according 
to the two
possible outputs $\sigma^\mu$. The $p$ patterns 
therefore generate a random partition of the coupling space into $2^p$ 
(possibly empty) cells
\beq\label{celldef}
C( \{ \sigma^\mu \}_{\mu=1,\ldots ,p} ) = 
\{ {\bf J};\;  \sigma^\mu = \mbox{sgn} ( {\bf J}\boldxi^\mu ) \;\forall \mu 
\}
\eeq
labeled by the $2^p$ output sequences 
$\boldsigma = \{ \sigma^\mu \}$ (fig.1).
\begin{figure}[htb]
\epsfysize=10cm
      \epsffile{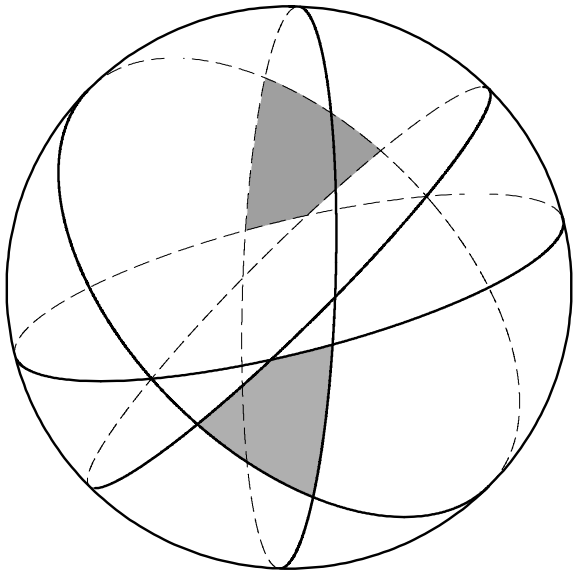}
\caption{
Symbolic representation of the random partition of a spherical coupling 
space for $N=3$ and $p=4$. The figure shows the existence of a
``mirror cell'' corresponding to the symmetry of (\ref{perc}) under
the transformation $({\bf J},\boldsigma) \mapsto (-{\bf J},-\boldsigma)$.}
\end{figure}

The relative cell size
$P(\boldsigma)=V(\boldsigma ) /$
$\sum_{{\mbox{\boldmath $\tau$}}}V({\mbox{\boldmath $\tau$} })$
gives the probability for generating the output ${\boldsigma}$ for
a given input sequence ${\mbox{\boldmath $\xi$}}^\mu$ with a coupling vector
${\bf J}$ drawn at random from a uniform distribution over the whole space
of couplings.
The natural scale of this quantity in the thermodynamic limit is
$\epsilon = 2^{-N}$. For the Ising perceptron this corresponds to a cell
containing just a single coupling vector. 
It is convenient to characterize the cell sizes by the 
{\it crowding index} 
$\alpha( \boldsigma )$ defined by
\beq
P( \boldsigma ) = \epsilon^{\alpha( \boldsigma)} \;.
\eeq
As discussed nicely in the Derrida part of \cite{DGP} the storage and
generalization properties of the perceptron are coded in the 
{\em distribution of cell sizes} defined by 
\beq
f(\alpha)=\lim_{N\to\infty} \frac{1}{N \log 2} \log 
    \sum_{\boldsigma}\delta(\alpha-\alpha(\boldsigma)) \;.
\eeq
To calculate this quantity within the framework of statistical mechanics one
uses the formal analogy of $f(\alpha)$ with the micro--canonical entropy of
the spin system $\boldsigma$ with Hamiltonian $\alpha(\boldsigma)$. 
It can hence be determined from the corresponding free energy
\beq\label{selfaver}
\tau (q) = - \lim_{N\to\infty} \frac{1}{N \log 2} \lla \log \sum_{\boldsigma}
           \exp(-q\log 2 \,\alpha(\boldsigma)) \rra 
= - \lim_{N\to 0} \frac{1}{N \log 2} \lla \log \sum_{\boldsigma}
       P^q (\boldsigma ) \rra 
\eeq
via Legendre-transformation with respect to the inverse temperature $q$
\beq\label{legendre}
f(\alpha) = \min_q [\alpha q - \tau(q)]\;.
\eeq
This procedure \cite{EW} is very similar to the so-called thermodynamic formalism in 
the
theory of multifractals \cite{FrPa,Halsey} where the multifractal spectrum
$f(\alpha)$ is introduced to characterize a probability measure by
the moments 
\beq
\langle P^q \rangle = \sum_{\boldsigma} P^q ( \boldsigma ) 
= \epsilon^{\tau(q)}\;.
\eeq
In this connection $\tau(q)$ is called the {\it mass exponent}. The only
new feature here is the additional average over the random inputs
$\boldsigma$ represented by $\lla\ldots \rra$ in (\ref{selfaver}).
The application of multifractal techniques to the theory of neural
networks was initiated by Monasson and O'Kane in their study of the
distribution of internal representations of multi-layer neural networks
\cite{MoKa}.

To perform the analysis for the perceptron we start with the 
definition of the cell size
\beq
P(\boldsigma) = \int d\mu({\bf J}) \prod_{\mu =1}^p 
\theta( \frac{1}{\sqrt{N}}
\sigma^\mu {\bf J} \boldxi^\mu)
\label{size}
\eeq
using the Heaviside step function $\theta(x)$. The integral measure 
$d\mu({\bf J})$
ensures that the total volume is normalized to 1. 

In the thermodynamic limit we expect both $\tau$ and $f$ to become 
self-averaging,
and we can therefore calculate the mass exponent (\ref{selfaver}) by
using the replica trick introducing $n$ identical
replicas numbered $a=1,\ldots,n$ to perform the average over the quenched 
patterns.
Moreover, we introduce a second
replica index $\alpha = 1,\ldots,q$ in order to represent the $q$-th power
of $P$ in (\ref{selfaver}) assuming as usual that the result can be
meaningfully continued to real values of $q$ \cite{PaVi}. 
Introducing integral representations 
for the  Heaviside function we arrive at a replicated partition function
given by
\begin{eqnarray}
\lla Z^n \rra & = & \lla  \sum_{\{ \sigma_\mu^a \} } \int\prod_{a,\alpha}
d\mu ( {\bf J}^{a,\alpha}) \; \prod_{\mu,a,\alpha}\theta(\sigma_\mu^a
\frac{1}{\sqrt{N}}\sum_j J_j^{a,\alpha} \xi_j^{\mu}) \rra \nonumber \\
 &=&\lla\sum_{\{ \sigma_\mu^a \} } \int \prod_{a,\alpha}
d\mu ( {\bf J}^{a,\alpha} ) \int_0^\infty \prod_{\mu,a,\alpha} 
\frac{d\lambda_\mu^{ a,\alpha}}
{\sqrt{2\pi}} \int \prod_{\mu,a,\alpha} \frac{dx_\mu^{
a,\alpha}}{\sqrt{2\pi}} \nonumber \\
 & & \exp\left\{ i \sum_{\mu,a,\alpha} 
 x_\mu^{a,\alpha} ( \lambda_\mu^{a\alpha} - \frac{1}{\sqrt{N}} \sigma_\mu^a
 \sum_i J_i^{a\alpha} \xi_i^\mu ) \right\} \rra \;.
\label{massgen}
\end{eqnarray}
The average over the quenched patterns $\boldxi^\mu$ can now be easily done.
To disentangle the remaining integrals we introduce the order
parameters 
\beq
Q_{ab}^{\alpha\beta} = \frac{1}{N} \sum_i J_i^{a\alpha} J_i^{b\beta}
\label{overlap}
\eeq
as the overlap of two coupling vectors, and their conjugates 
$\hat{Q}_{ab}^{\alpha\beta}$.
The spherical as well as the Ising constraints restrict
the self-overlap $Q_{aa}^{\alpha\alpha}$ to 1. The other values of the order
parameter matrices are
obviously invariant under simultaneous commutations of $a\leftrightarrow b$
and $\alpha \leftrightarrow \beta$. We then find
\begin{eqnarray}
\lla Z^n \rra &=& \int \prod_{(a,\alpha)<(b,\beta)} 
\frac{dQ_{a,b}^{\alpha,\beta}
\, d\hat{Q}_{a,b}^{\alpha,\beta}}{2\pi / N} \nonumber \\
 & & \exp\left\{ N \left[ 
-\sum_{(a,\alpha)<(b,\beta)}Q_{a,b}^{\alpha,\beta} 
\hat{Q}_{a,b}^{\alpha,\beta}
+\gamma \log G_0(Q_{a,b}^{\alpha,\beta})
+ \log G_1(\hat{Q}_{a,b}^{\alpha,\beta}) \right] \right\}
\label{int}
\end{eqnarray}
with
\begin{eqnarray}
G_0 (Q_{ab}^{\alpha\beta}) & = &\int_0^{\infty} \prod_{a,\alpha} 
\frac{d\lambda^{a,\alpha}}{\sqrt{2\pi}} \int \prod_{a,\alpha} 
\frac{dx^{a,\alpha}}{\sqrt{2\pi}} \sum_{\{ \sigma^a \} }
\exp{ \left\{ i \sum_{a,\alpha} x^{a\alpha}\lambda^{a\alpha}\sigma^{a}
       -\frac{1}{2} \sum_{a,b,\alpha,\beta} x^{a\alpha}x^{b\beta}
       Q_{ab}^{\alpha\beta} \right\} }\nonumber\\
G_1 (\hat{Q}_{a,b}^{\alpha,\beta}) &=& \left[ \int\prod_{a,\alpha}
d\mu({\bf J}^{a\alpha}) \exp\left\{ \sum_{(a,\alpha)<(b,\beta)} 
\hat{Q}_{a,b}^{\alpha,\beta} \sum_i J_i^{a\alpha} J_i^{b\beta} \right\}
\right]^{1/N} \;
\label{Gnull}
\end{eqnarray}
 $(a,\alpha)<(b,\beta)$ denotes either $a<b$ or $a=b,\alpha<\beta$
and counts the elements above the main diagonal in the order parameter
matrices. The integrals over the order parameters in (\ref{int}) can be done
using the saddle point method.

To find the correct saddle point is in general a very difficult task.
A simple ansatz is the replica symmetric one. For the present situation
it is important to note that the output sequences $\{ \sigma_\mu^a \}$ carry 
only one replica index. The typical overlap of two coupling vectors within
one cell (same output sequence $\{ \sigma_\mu^a \}$) will hence 
in general be different
from the typical overlap between two coupling vectors belonging to different
cells (different output sequence $\{ \sigma_\mu^a \}$).
Therefore we have to introduce already within the replica symmetric (RS) 
approximation two
different overlap values in order to determine the saddle point
of (\ref{int})( see \cite{MoKa}):
\beq
Q_{ab}^{\alpha\beta} = \left\{
\bay{ll}
1 & \mbox{if} \;\; (a\alpha)=(b\beta) \\
Q_1 & \mbox{if} \;\; a=b,\;\alpha\neq\beta \\
Q_0 & \mbox{if} \;\; a\neq b
\eay \right. \;.
\label{QRS}
\eeq
In accordance with the above discussion $Q_1$ then denotes the typical
overlap {\em within
one cell}, whereas $Q_0$ denotes the overlap {\em between different cells}. 
The structure of the conjugated order parameter is analogous, having 
non-unite diagonal elements $\hat{Q}_2$.

Plugging this RS ansatz into (\ref{massgen}) one
realizes that  $Q_0 = \hat{Q}_0 = 0$ always solves the saddle point equations
for $Q_0$ and $\hat{Q}_0$. This has an obvious
physical interpretation: Due to the symmetry of (\ref{perc}) and therefore of
the crowding index $\alpha(\boldsigma)$ under the transformation 
$({\bf J}, \boldsigma ) \leftrightarrow (-{\bf J}, -\boldsigma )$ every cell 
has a ``mirror cell'' of same size and shape on the ``opposite side'' of the 
coupling  space (see fig.1). $Q_0=0$ simply reflects this symmetry. 
It can be explicitly broken by introducing a threshold in (\ref{perc}). Note 
that $Q_0 = \hat{Q}_0 = 0$ means formally that the quenched average over the
input patterns can be performed as an {\em annealed} average.

\section{The spherical perceptron}
\subsection{Replica symmetry}

In the case of the spherical perceptron the coupling space is restricted to
the N-dimensional hyper--sphere defined by the global spherical constraint
${\bf J}^2 = N$. In the large-$N$ limit this gives rise to the 
integral measure
\beq
d\mu({\bf J}) = \prod_i \frac{dJ_i}{\sqrt{2\pi e}} \; \delta(N-{\bf J}^2) .
\eeq
Using the replica symmetric ansatz (\ref{QRS}) with $Q_0 = \hat{Q}_0 = 0$
leads to the mass exponent
\begin{eqnarray}
  \label{massexp}
  \tau(q) = -\frac{1}{\log  2} \;\mbox{extr}_{Q_1,\hat{Q}_{1,2}}&& \left[
            \frac{q}{2}(\hat{Q}_2-1) + \frac{q(q-1)}{2}\hat{Q}_1 Q_1
            -\frac{1}{2}\log(\hat{Q}_2+(q-1)\hat{Q}_1)
            -\frac{q-1}{2}\log(\hat{Q}_2-\hat{Q}_1)\right. \nonumber\\
 &&\left.     +\gamma \log \; 2\int Dt H^q\left( \sqrt{ \frac{Q_1}{1-Q_1} } t 
            \right)\right],
\end{eqnarray}
where we introduced the abbreviations $Dt = dt\; \exp(-t^2/2) / \sqrt{2\pi}$
for the Gaussian measure and $H(x) = \int_x^{\infty} Dt$.  
As is well known for spherical models, the saddle point equations for the
conjugated order parameters $\hat{Q}_{ab}^{\alpha\beta}$ can be solved 
explicitly which in the present case yields 
\beq
\tau(q) = - \frac{1}{\log 2}  \;\mbox{extr}_{Q_{1}}\left[ \frac{1}{2}\log (1+(q-1)Q_1) + 
\frac{q-1}{2} \log ( 1-Q_1 ) + \gamma \log \; 2 \int Dt H^q\left( 
\sqrt{ \frac{Q_1}{1-Q_1} } t \right)\right].
\label{rsmass}
\eeq
The order parameter $Q_1$ is 
self-consistently determined by the saddle point equation
\beq
\frac{Q_1}{1+(q-1)Q_1} = \frac{\gamma}{2\pi} 
\frac{\int Dt H^{q-2}\left( \sqrt{ \frac{Q_1}{1-Q_1} } t \right)
\exp \left\{ -\frac{Q_1}{1-Q_1} t^2 \right\} }
{\int Dt H^{q}\left( \sqrt{ \frac{Q_1}{1-Q_1} } t \right) } \;.
\label{saddle}
\eeq
The multifractal spectrum $f(\alpha)$ resulting from a
numerical solution of these equations is shown in fig.2 for various
loadings $\gamma$. 

\begin{figure}[htb]
 \epsfysize=10cm
      \epsffile{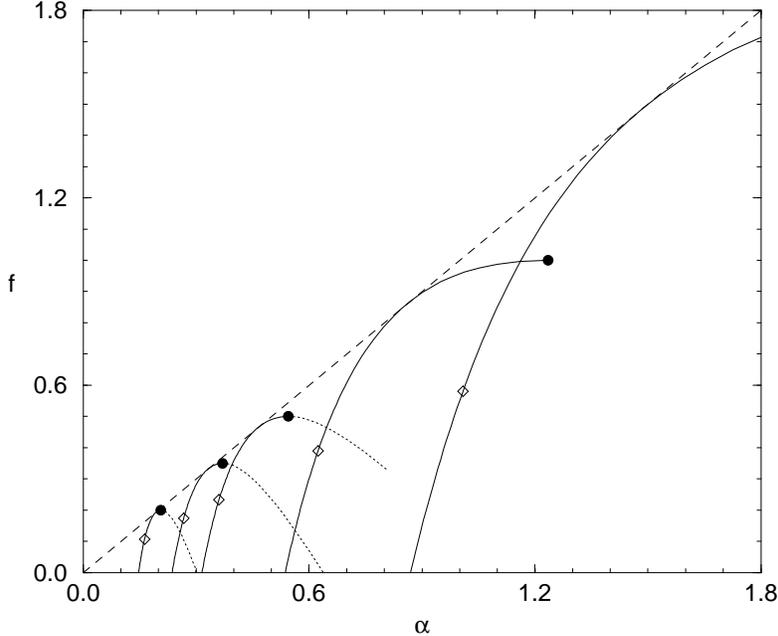}
\caption
{Multifractal spectrum $f(\alpha)$ characterizing the cell structure of
the coupling space of the spherical perceptron for various values
of the loading parameter $\gamma=0.2,0.35,0.5,1.0,2.0$ (from left to right).
The curves end at their maxima because of the divergence of the mass
exponent $\tau(q)$ for negative $q$ (corresponding to the dotted parts).
Replica symmetry holds between the diamonds and the maxima.}
\end{figure}

For small values of $\gamma$ we find the typical bell-shaped form 
of $f(\alpha)$. The zeros $\alpha_{min} (\gamma)$ specify the RS estimate of
the largest cell occurring with non-zero probability \cite{BiehlOpper}.

The most frequent cell size corresponds to the maximum of $f(\alpha)$ and is
therefore given by $\alpha_0(\gamma)= \mbox{argmax}(f(\alpha))$. 
For large $N$ cells of this size dominate
the total number of cells exponentially, i.e. a randomly chosen {\em output
sequence} $\boldsigma$ will be found with probability 1 in a cell of size 
 $\alpha_0$. Hence $\epsilon^{\alpha_0(\gamma)}$ (cf. (3))
is the {\em typical} volume of
couplings realizing $p=\gamma N$ random input-output mappings as determined
within a standard Gardner calculation \cite{gardner}. This volume becomes
zero, i.e. $\alpha_0(\gamma)\to\infty$, for $\gamma\to 2$ (cf. fig.1) in
accordance with the Gardner result \cite{gardner}.

In addition we can infer from
$f(\alpha_0(\gamma))$ the {\em typical number} of cells as first 
calculated with the help of geometrical methods by Cover \cite{cover}.
For small loading ratios $\gamma$
we find  $f(\alpha_0) = \gamma$, i.e. all $2^{\gamma N}$ possible cells (or
almost all of them) do indeed occur. The storage problem for these
 $\gamma$ values
is then solvable with probability one. For $\gamma > 2$ we have
$f(\alpha_0)=f(\infty)<\gamma$ implying that only an exponentially small
fraction of all possible cells can be realized. It is then typically
impossible to find couplings realizing a randomly generated set of 
input-output mappings. 
The multifractal analysis of
the coupling space hence nicely reconciles the previously complementary
approaches to the storage problem of the perceptron by Cover and Gardner
respectively. From both the analysis of $\alpha_0(\gamma)$ and of 
$f(\alpha_0(\gamma))$ one finds the well known result $\alpha_c=2$. 

Although the cells with volume $\alpha_0$ are the most frequent ones their
joint contribution to the total {\em volume} of the sphere is negligible.
Since 
\beq
1=\sum_{\boldsigma} P(\boldsigma)=\int_0^{\infty} d\alpha
    \exp(N[f(\alpha)-\alpha])
\eeq
a saddle point argument reveals that the cells with size $\alpha_1(\gamma)$
defined by $f'(\alpha_1)=1$ dominate the volume. Cells of larger size are
too rare, those more frequent are too small to compete. Consequently a
randomly chosen {\em coupling vector} ${\bf J}$ will belong with probability 1
to a cell of size $\alpha_1$. By the definition (\ref{celldef}) 
of the cells
all other couplings of this cell will give the same output for all patterns
 $\boldxi^{\mu}$. Therefore $\epsilon^{\alpha_1(\gamma)}$ is nothing but
the volume of the version space of a teacher perceptron chosen at random
from a  uniform probability distribution 
on the sphere of possible perceptrons. From it (or equivalently
from $Q_1(q=1,\gamma))$ one can determine the generalization error as a
function of the training set size $\gamma$ reproducing the results of
\cite{GT}. 

The main properties of the perceptron can hence be derived from the
multifractal spectrum $f(\alpha)$ of the cell size distribution in the
coupling space. Below we show that the RS ansatz together with the
assumption $Q_0=0$ gives valid results for $0\leq q\leq 1$.

There is also a
close formal analogy between the calculation of $f(\alpha)$ and the standard
Gardner approach with $q$ playing the role of the replica number in the
Gardner calculation. Since from (\ref{legendre}) we have $q=df/d\alpha$
the calculation of $\alpha_0$ is related to $q\to 0$ whereas the
generalization problem concentrating on $\alpha_1$ corresponds to $q\to 1$. 
These limits  of the replica number in Gardner calculations are well known
to correspond to the storage and generalization
problems respectively \cite{OH}.

\subsection{Longitudinal instability of replica symmetry}

The results of the previous paragraph were obtained within the RS
approximation and using $Q_0=0$. The discussion of their validity
requires a careful determination of the stability of these ansatzes for the
different values of $q$.
We first discuss the stability with respect to longitudinal fluctuations in
$Q_0$, i.e. we search for a RS 
saddle point solution where the symmetry giving rise to $Q_0 =0$ is 
spontaneously broken. The full replica symmetric saddle point equations are of the form 
\begin{eqnarray}
0 &=& \frac{Q_0}{(1+(q-1)Q_1-Q_0)^2} \nonumber \\
 & &  -\frac{\gamma}{2\pi(1-Q_1)} \int Dy
\left( \frac{\int Dt ( H_+^{q-1}-H_-^{q-1} ) \exp \left\{ -
\frac{(\sqrt{Q_0}y+\sqrt{Q_1-Q_0}t)^2}{2(1-Q_1)} \right\} }
{\int Dt ( H_+^q + H_-^q ) } \right)^2 \;, \\
0 &=& \frac{Q_1-Q_0}{1+(q-1)Q_1-Q_0} + \frac{Q_0(1-Q_1)}{(1+(q-1)Q_1-Q_0)^2}
\nonumber \\
 & & - \frac{\gamma}{2\pi} \int Dy
\frac{\int Dt ( H_+^{q-2}+H_-^{q-2} ) \exp \left\{ -
\frac{(\sqrt{Q_0}y+\sqrt{Q_1-Q_0}t)^2}{(1-Q_1)} \right\} }
{\int Dt ( H_+^q + H_-^q ) } 
\end{eqnarray}
where 
\beq
H_\pm = H\left( \pm\frac{\sqrt{Q_0}y+\sqrt{Q_1-Q_0}t}{\sqrt{1-Q_1}}\right) \;.
\eeq
Linearizing these equations in $Q_0$ we find that new solutions $Q_0>0$
bifurcate continuously from $Q_0=0$ at 
\beq\label{linear}
q_\pm = 1 \pm \frac{\sqrt{\gamma}}{Q_1(\gamma)}
\label{qc}
\eeq
From (\ref{linear}) and (\ref{saddle}) the two
transition points $q_\pm(\gamma)$ for positive/negative inverse temperature
$q$ can be determined explicitly.
In the range $q_- < q < q_+$, there exists only the solution with $Q_0=0$
which becomes unstable at the transition points.

\subsection{Divergence of negative moments}
The expression (\ref{massexp}) for the replica symmetric mass exponent
has been calculated for positive integer $q$. The
continuation to negative values of $q$ gives rise to  divergences at
\beq
\delta = \frac{1+(q-1)Q_1}{1-Q_1} = 0
\label{delta}
\eeq
as can be realized from an asymptotic analysis
of the integrand in the last term of (\ref{massexp}). Because of
$H(t) \propto \exp(-t^2/2)/\sqrt{2\pi}t$
for large $t$ we get an exponential part of this term proportional to
$\exp(-\delta t^2 /2)$ and the whole integral converges
only if $\delta > 0$. For $\delta\to 0$ we therefore find that $\tau$ tends
to $-\infty $. The global minimum of (\ref{massexp}) with respect to $Q_1$
is hence no longer given
by the saddle point described by (\ref{rsmass},\ref{saddle})
which realizes only a local minimum with respect to $Q_1$. There is hence a
{\em discontinuous} longitudinal transition at $q=0$ with $Q_1$ jumping from
the solution of (\ref{saddle}) to $1/(1-q)$. 
As a consequence $f(\alpha)$ is not defined for $f'(\alpha)<0$
and the curves for
$f(\alpha)$ as obtained above by using the saddle-point equations  are 
reliable only for $q\geq 0$, i.e. for
positive slope. The parts corresponding to $q<0$ are 
dotted in fig.2. 

It is tempting to speculate that the observed divergence for negative $q$ is
due to the existence of {\it empty cells} $V(\boldsigma)=0$ in
(\ref{selfaver}). In the theory of neural
networks (and more generally of classifier systems) the possibility of
output sequences impossible to implement by the system is related to the
Vapnik-Chervonenkis- (VC-) dimension $d_{VC}$
of the class of networks under consideration \cite{VC,vcrev}. It has been
notoriously difficult to determine the VC-dimension of a neural network 
from statistical
mechanics calculations since the definition of the VC-dimension involves a
{\em supremum} over all possible pattern sets rather than the {\em average} 
featuring in
(\ref{selfaver}). The above analysis of the instability with
respect to $Q_1,\hat{Q_1}$ for $q<0$ reveals that the multifractal formalism
in the present form is unfortunately also unable to determine $d_{VC}$ since 
the
divergence of negative moments of $V(\boldsigma)$ occurs for {\it all}
values of $\gamma$. The pattern average performed in (\ref{selfaver}) does not
allow to decide whether the observed divergence of $\tau$ is due to the fact
that $\gamma>d_{VC}/N$ or is due to exceptional pattern realizations that
give rise to empty cells also if $\gamma<d_{VC}/N$ \cite{rem2}. Note in this
connection also that similar divergences in the theory of multifractals
\cite{BoTe} are related to cells with volume which is non-zero but 
decreases for
$N\to\infty$ {\it quicker} then exponentially. For the perceptron on the
other hand it is known that empty cells exist for all values of $N$
\cite{cover}.
\subsection{Transversal instability of replica symmetry and percolation}
In addition to the longitudinal instabilities discussed in the previous
paragraph there is the possibility of a transversal instability invalidating
the RS ansatz \cite{AT} which we study now. The replica symmetric ansatz
(\ref{QRS}) is formally similar to a 
one-step replica symmetry broken solution (1RSB) of a  standard 
replica calculations \cite{MPV}.
It is advantageous to use the results and the notation given in \cite{TDK}.
After some lengthy calculations we arrive at 4 different eigenvalues
corresponding to the replicon modes denoted by (0,1,1), (1,2,2), (0,2,2)
and (0,2,1) in \cite{TDK}. We give here only the result for the
(0,1,1)-mode which is found to be the first to become unstable.
\beq
\lambda(0,1,1) = \frac{1- \frac{1}{\gamma}(q-1)^2 Q_1^2}
{(1+(q-1)Q_1)^2} \;.
\label{ew}
\eeq
It vanishes exactly at the two points calculated in (\ref{qc}) describing
the instability with respect to longitudinal $Q_0$-fluctuations. 
Similar to the SK-model \cite{SK} in zero field the
longitudinal and transversal instability of the RS-solution occur hence for
the same temperature $1/q$ \cite{rem1}. Due to the
divergences for $q<0$ only
$q_+$ is of further relevance. The AT-points are therefore determined by 
$q_+ (\gamma)$ which are marked in fig.2 by the diamonds.

The eigenvalue $\lambda(0,1,1)$ describes fluctuations in that part of the overlap matrix 
having only $Q_0$-entries, i.e. corresponding to the overlaps between
different cells. This is reasonable since for the spherical
perceptron the cells themselves are known to be convex. No RSB is hence
expected to be necessary to
describe the structure of a {\em single} cell \cite{gardner}. The
instability to RSB found above concerns the distribution of overlaps
between different cells which must now be characterized by two parameters.
The smaller one remains equal to zero reflecting still the symmetry of 
(\ref{perc}). The other one is larger than zero and describes the 
formation of clusters made of cells of identical size.

In order to interpret the RSB transition in physical terms we allude again
to the analogy with the SK-model. There the analogous instability corresponds to broken
ergodicity, i.e. to the fact that not all parts of the phase space 
can be reached from a given initial condition. In the perceptron problem single
spin flips in the output sequence
$\boldsigma$ are equivalent to hops between neighboring cells in the
coupling space. The breakdown of $Q_0=0$ at $q_+$ hence signals 
that starting in a cell of a size corresponding to $q_+$, i.e. starting 
with a spin configuration $\boldsigma$ with energy $\alpha(\boldsigma)$ corresponding to
$q_+$ it becomes impossible to reach the ``mirror cell'' by hops using only cells of the 
same or larger sizes, i.e. via spin configurations 
$\boldsigma$ with the same or smaller energy. Since the relative number of 
larger cells is exponentially small we can interpret the observed breaking of
RS as a {\it percolation transition} in the infinite dimensional space of
couplings. For $0<q<q_+$ the cells of size $\epsilon^{\alpha(q)}$
percolate in coupling space in the sense that they can all be reached 
from each other by entering only
cells of the same size. For $q>q_+$ this is no longer true and the cells
form clusters isolated from each other.

\section{The Ising perceptron}
\subsection{Replica symmetry}

In the Ising perceptron the entries of the coupling vector are restricted
to $J_i = \pm 1,\;i=1,\ldots,N$, the full coupling space is hence 
given by the $2^N$
corners of a $N$-dimensional hypercube. The cells are therefore represented
by discrete sets,
their probability measure is given by the number of elements multiplied with
$2^{-N}$. Thus, the coupling space measure is to be modified according to
\beq\label{isingsize}
\int d\mu({\bf J}) \mapsto 2^{-N} \sum_{\bf J} \;.
\eeq
Following the general procedure described in the section II we arrive at
\beq \label{isgen}
\tau(q)= q + \lim_{n\to 0} \frac{1}{n \log 2} \;\mbox{extr}_{(Q_{ab}^{\alpha\beta}),
(\hat{Q}_{ab}^{\alpha\beta})} \left( \frac{1}{2} \sum_{(a\alpha)\neq(b\beta)}
\hat{Q}_{ab}^{\alpha\beta} Q_{ab}^{\alpha\beta} - 
\log G_1((\hat{Q}_{ab}^{\alpha\beta}))
- \gamma \log G_0((Q_{ab}^{\alpha\beta})) \right)
\eeq
with
\beq
G_0((Q_{ab}^{\alpha\beta}))  = 
\sum_{\sigma^a} \int_0^\infty \prod_{a\alpha} d\lambda^{a\alpha}
\int \prod_{a\alpha} \frac{dx^{a\alpha}}{2\pi} \exp \left\{
i\sum_{a\alpha} x^{a\alpha} \lambda^{a\alpha} \sigma^a 
-\frac{1}{2}\sum_{ab\alpha\beta} x^{a\alpha}x^{b\beta}Q_{ab}^{\alpha\beta}
\right\}
\eeq
and
\beq
G_1((\hat{Q}_{ab}^{\alpha\beta})) = \sum_{J^{a\alpha}} \exp \left\{
\frac{1}{2} \sum_{(a\alpha)\neq(b\beta)} \hat{Q}_{ab}^{\alpha\beta}
J^{a\alpha} J^{b\beta} \right\} \; .
\eeq

Contrary to the spherical case the conjugated parameters
$\hat{Q}^{\alpha\beta}$ can not be eliminated analytically. 
The replica symmetric expressions are obtained by introducing 
the saddle point
structure (\ref{QRS}) for both the overlaps and their conjugates.  
Starting again with the solution  $Q_0=\hat{Q}_0=0$  
we get for the mass exponent
\begin{eqnarray}
\tau(q) &= \frac{1}{\log 2} \;\mbox{extr}_{Q_1,\hat{Q}_1}&\left[
\frac{q \hat{Q}_1}{2}(1+(q-1)Q_1) -\log \int Dt \; \cosh^q(\sqrt{\hat{Q}_1}t)
\right. \nonumber \\
 & & \left.
-\gamma \log \,2\int Dt\, H^q\left(\sqrt{\frac{Q_1}{1-Q_1}} t\right)\right]\;.
\label{massising}
\end{eqnarray}
The extremization in this equation is again somewhat subtle \cite{Cocco}.
There are two, qualitatively different 
possibilities: 
$(i)$ The first one is given by  $Q_1=1$
and $\hat{Q}_1=\infty$ and  hence lies at the boundary of allowed
values of the saddle-point parameters. It can be studied analytically and
leads to $\tau(q)=q-1$. The corresponding multifractal spectrum is given by
a single point $f=\alpha=1$. No dependence on $\gamma $ remains. 
The value $\alpha=1$ describes cells containing just a single coupling
vector in accordance with $Q_1=1$. Their total number is of order 
$\epsilon^{-f}=2^N$, and hence they form a
macroscopic part of the cell number  as well as  of the total 
coupling space volume. Since the total number of
cells is $2^{\gamma N}$ this solution can exist for $\gamma\geq1$
only.
\noindent
$(ii)$ The second solution solves the saddle-point equations
\begin{eqnarray}\label{sadising}
Q_1 & = & \frac{ \int Dt \;\cosh^{q-2}(\sqrt{\hat{Q}_1}t)
\, \sinh^2(\sqrt{\hat{Q}_1}t)}
{\int Dt \; \cosh^q(\sqrt{\hat{Q}_1}t)} \nonumber\\
\hat{Q}_1 & = & \frac{\gamma}{2\pi(1-Q_1)}
\frac{\int dt \, \exp\left\{ -\frac{(1+Q_1)t^2}{2(1-Q_1)}\right\}  
H^{q-2}\left(\sqrt{\frac{Q_1}{1-Q_1}} t\right)}
{\int Dt\, H^q\left(\sqrt{\frac{Q_1}{1-Q_1}} t\right)}\;.
\end{eqnarray}
It lies inside the intervals for the parameters and is to be
determined numerically. It exists for all $q$ if $\gamma\leq 1$
and disappears for fixed $\gamma>1$ at a sufficiently negative $q$.

A numerical comparison of the corresponding local maximums/minimums of 
$\tau$  leads to the following scenario: For $\gamma<1$ solution $(ii)$
always gives the global extremum. For $\gamma>1$ this is only the case 
for $q>q_{disc}(\gamma)$. At this threshold the extremum $(i)$ 
becomes the global one by a discontinuous transition. The occurrence of such
transitions is the trademark of 
neural networks with Ising couplings \cite{KM,Cocco}. The
smooth curves of $f(\alpha)$ then terminate and are completed by a single 
point at (1,1) (see fig.3). Hence for  $\gamma>1$ the spectrum is nonzero 
only in a certain region $\alpha_{min}<\alpha<\alpha_{max} <1$ and 
at the isolated point at $\alpha=1$.

\begin{figure}[htb]
 \epsfysize=10cm
       \epsffile{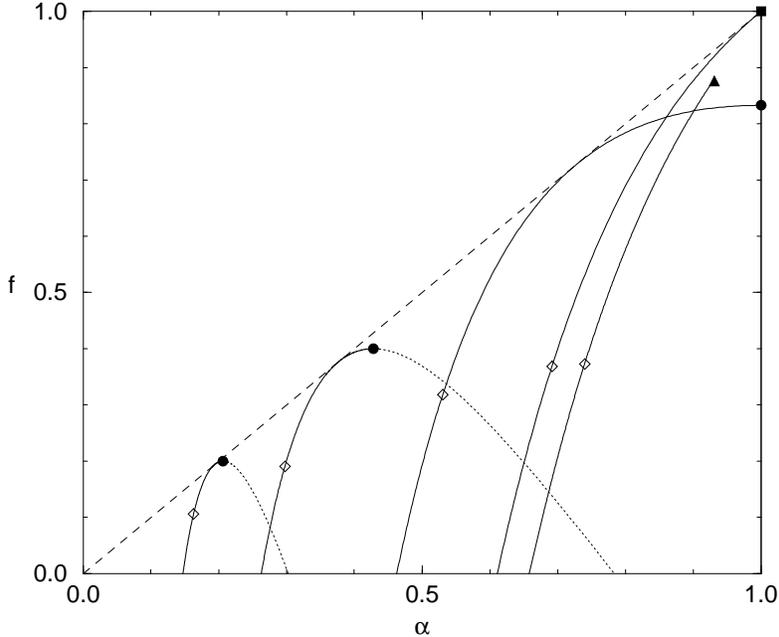}
\caption{
Multifractal spectrum $f(\alpha)$ describing the cell structure of the
coupling space of the Ising perceptron with loading ratios
$\gamma=0.2,0.4,0.833,1.245,1.4$ (from left to right).
The curves end at their maxima due to the divergence of the mass exponent
$\tau(q)$ for negative $q$. Between the diamonds and the maxima RS holds. 
The triangle denotes the discontinuous transition to $Q_1=1$. The isolated point (1,1) 
is marked by the square.}
\end{figure}

The general form of the multifractal spectrum resembles that of the spherical perceptron.
In fact, for $\gamma\lesssim 0.25$ the curves almost coincide. This was to
be expected because the cells are still relatively large and do not ``feel''
the discreteness of the Ising couplings. For larger $\gamma$, however, it becomes decisive
that in the Ising perceptron the cell sizes are
restricted to $\alpha\leq 1$. All values $\alpha>1$ correspond 
to empty cells.

The storage capacity is therefore not given by $\alpha_0\to\infty$ as in the
case of the spherical perceptron but is determined by 
$\alpha_0(\gamma) = \mbox{argmax}(f(\alpha,\gamma)) = 1$.
As can be seen from fig. 3 this holds 
for $\gamma_c=0.833$ the well known result obtained in \cite{KM}. 
In fact $\alpha_0(\gamma)=1$ is
equivalent to the zero-entropy condition frequently used for neural networks
with discrete couplings. Let us define the entropy
\beq
s=\lim_{N\to\infty}\frac{1}{N} \lla \log \cal{N}\rra
\eeq
where $\cal{N}$ denotes the number of couplings that can implement a mapping
between $\gamma N$ random inputs {\em and} outputs. Therefore $\cal{N}$ is
related to the size of the typical cell and from
eqs.(\ref{size},\ref{isingsize}) one finds $s=(1-\alpha_0)\log 2$. Hence
$s=0$ is equivalent to $\alpha_0=1$.

Similarly the generalization behaviour of the Ising perceptron differs from
that of the spherical one. There is a well--known
{\em discontinuous transition to perfect generalization} 
at $\gamma_g=1.245$ \cite{Gyo}. It shows up in the
multifractal spectrum $f(\alpha,\gamma)$ as the point where
$\alpha_1(\gamma_g)=1$ with $\alpha_1$ again defined by $f'(\alpha_1)=1$. 
At this value of the loading parameter the discontinuous
transition to $Q_1=1$ occurs and the coupling
space becomes dominated by cells with exactly one element. This is the
intuitive reason for the transition to perfect generalization; there is only
one coupling vector left that performs perfectly on the training set: the
teacher herself.

\subsection{Continuous replica symmetry breaking and percolation}

Similar to section 2 we have analyzed the longitudinal and transversal
stability of the RS solution with $Q_0=\hat{Q}_0=0$ 
by calculating the relevant eigenvalues of the fluctuation
matrix. The results are qualitatively the same. 
Using the linearization of the complete replica
symmetric saddle point equations at $Q_0=\hat{Q}_0= 0$ we find again a new
solution with $Q_0, \hat{Q}_0 > 0$ emerging continuously at the two values
$q_\pm$ satisfying 
\beq\label{evl}
\sqrt{\gamma}=\pm\hat{Q}_1 (q_\pm -1)(1-Q_1)(1+(q_\pm-1)Q_1)
\label{isingrsb}
\eeq

The analysis of the transversal fluctuations reveals that the first mode to
become unstable is again the replicon mode (0,1,1). Due to the 
existence of two order parameter matrices the analysis is now more
involved. Following the argumentation in \cite{gardner} we first
determine the eigenvalues in the two building blocks of the fluctuation
matrix and find for the 
(0,1,1)-eigenvalue of the overlap fluctuations 
\beq\label{ev1}
\lambda^{(Q)}(0,1,1)=-\gamma^{-1}(q-1)^2(1-Q_1)^2\hat{Q}_1^2,
\eeq
and for  the conjugated matrix
\beq\label{ev2}
\lambda^{(\hat{Q})}(0,1,1)=-(1+(q-1)Q_1)^2.
\eeq
The total fluctuation matrix for this replicon mode is then given by 
\beq\label{ev3}
\left(
\bay{cc}
\lambda^{(Q)}(0,1,1) & 1 \\
1 & \lambda^{(\hat{Q})}(0,1,1)
\eay
\right)\;.
\eeq
and the breakdown of RS is signaled by a change of the sign of its 
determinant. 
From (\ref{ev1}),(\ref{ev2}),and (\ref{ev3}) we find that the
local transversal instability occurs again at the same values $q_\pm(\gamma)$ 
given by (\ref{evl}) for which the RS
solution with $Q_0=\hat{Q}_0= 0$ becomes longitudinally unstable. 

In fig.3 the breakdown of the local stability of RS is again marked 
by the diamonds.
Inside the interval $(\alpha(q_+),\alpha_0)$ RS is locally stable.
As discussed already for the spherical case, in this region the solutions are 
symmetric under the reflection symmetry of the original system (\ref{perc})
and the cells of every fixed crowding index 
$\alpha\in (\alpha(q_+),\alpha_0)$ percolate in coupling space.
Outside this interval the overlaps between
different cells must again be described by two (or more) order parameters. 
The smaller one
vanishes, and still reflects the symmetry of (\ref{perc}). The other
one takes a value in $(0,Q_1)$ with $Q_1$ being the overlap within
one single cell and describes the size of the connected clusters remaining
below the percolation threshold.

\subsection{Divergence of negative moments}
For $q<0$ there occurs an analogous divergence of $\tau$ for small $\delta$
defined by (\ref{delta}) as in the spherical case which gives rise to a similar 
discontinuous transition with respect to $Q_1$. 
As a result the $f(\alpha)$--curves do not continue into 
regions of negative slope for {\it{any}} $\gamma$. It is hence
again impossible to infer the (still unknown \cite{stephan}) VC-dimension of
the Ising-perceptron from the multifractal analysis. Note that the only value
that could in principle be obtained from a statistical mechanics analysis is
what is called the {\it typical} VC-dimension that gives the maximal pattern
set size for which {\it typically} no empty cells occur \cite{MeEn}. Numerical
investigations suggest that this value is equal to $N/2$ \cite{Stambke,MeEn}.

\subsection{Discontinuous replica symmetry breaking}
Contrary to the case of the spherical perceptron there is an inconsistency
even within the region of {\em local} stability of the RS ansatz. For
$.833<\gamma<1.245$ the multifractal spectrum $f(\alpha)$ continues to values $\alpha >1$ 
corresponding to the unphysical region of cells having less then one but more 
then zero elements. 

We therefore expect a discontinuous transition to RSB already in the region
of local stability of RS. Due to the fact that a single cell is not necessarily
connected for the Ising perceptron, this transition is likely to take
place inside the blocks describing the overlaps {\it within a cell}. 
The global reflection symmetry remains unbroken and
therefore the typical overlaps between two cells stay zero. 
We can hence calculate the annealed average $\lla Z \rra$ of the partition
function.
After a standard calculation we find the following one--step RSB (1RSB)
result for the mass exponent 
\begin{eqnarray}
  \label{taursb}
\tau^{1RSB}(q)&=& \frac{1}{\log 2} \mbox{extr}_{m,Q_1,Q_2,\hat{Q}_1,\hat{Q}_2}
\left[ \frac{1}{2} q(q-m)Q_1\hat{Q}_1 + \frac{q}{2}\hat{Q}_2(1+(m-1)Q_2)
\right.\nonumber\\
 & &-\log \int Dz_1 \left[ \int Dz_2 
\cosh^m(\sqrt{\hat{Q}_1}z_1+\sqrt{\hat{Q}_1
-\hat{Q}_2} z_2) \right]^{\frac{q}{m}} \\
 & & \left. - \gamma \log 2 \int Dt_1 \left[ \int Dt_2
H^m ( \frac{ \sqrt{Q_1} t_1 + \sqrt{Q_2-Q_1} }{\sqrt{1-Q_2}} ) 
\right]^{\frac{q}{m}}
\right]\nonumber
\end{eqnarray}
where $Q_2$ is the order parameter inside the $m\times m$ diagonal
blocks of the overlap matrix and  $Q_1$ is the  entry outside these
blocks. $\hat{Q}_{1,2}$ are the corresponding conjugated quantities. A
similar expression was obtained in \cite{Cocco}.

Guided by previous experience \cite{KM} we look for an extremum with 
$Q_2=1$ and $\hat{Q}_2=\infty $.  One then finds 
\begin{equation}
  \label{tauq21}
  \tau^{1RSB}(q) = \mbox{extr}_m \left[ q(1-\frac{1}{m}) + \tau^{RS}(
  \frac{q}{m} ) \right]
\end{equation}
where we have used the RS mass exponent $\tau^{RS}$ from (\ref{massising}).
The saddle point equation with respect to $m$ is then given by 
\begin{equation}
  \label{abbruch}
  1= \frac{d\tau^{RS}}{dq} (\frac{q}{m}) = \alpha^{RS}.
\end{equation}
Comparing the values of $\tau^{RS}$ and $\tau^{1RSB}$ one finds that for
$\gamma >.833$ there is indeed a {\it discontinuous transition}  to this 1RSB solution when
$\alpha$ gets larger than $1$. Crossing this point $\tau(q)$ becomes
proportional to $q$ implying that $f(\alpha)$ stops at $\alpha=1$. This
removes inconsistency noted above. For $.833 < \gamma <1.245$ the
cells contributing most to the total volume (the $\alpha_1$-cells) contain
exponentially many couplings. The majority of cells,
however, comprise only sub-exponentially many couplings. This situation is
correctly described by the 1RSB solution with $Q_1 <1$ characterizing the
$\alpha_1$-cells and $Q_2=1$ characterizing the typical ones. With increasing $\gamma$
the $\alpha_1$-cells shrink and the typical cells disappear. At
$\gamma=1.245$ we have $\alpha_1=1$ and correspondingly $Q_1=1$. At this
point $\tau$ is again given by the {\it minimum} in (\ref{isgen}) since
$q>1$. Therefore the 1RSB solution has to be rejected and the discontinuous
transition disappears. The 1RSB solution also becomes intrinsically
inconsistent since there is ``no room left'' for a $Q_2$ with $Q_1<Q_2\leq 1$.
For $\gamma>1.245$ the 1RSB solution finds its natural continuation in the
RS solution with $Q_1=1$ that gives rise to 
the gap in the $f(\alpha)$ spectrum as discussed
in the first paragraph of this section.

\subsection{Numerical results}

For the Ising perceptron the phase space is discrete and the analytical
results discussed above can be checked by numerical enumerations over all the
possible $2^N$ coupling vectors $J_i=\pm 1$. Although these techniques
are naturally confined to rather low values of $N$ it is interesting to see
whether the asymptotic behaviour already shows up in small samples. We have
performed enumerations for values of $N$ between 10 and 30 according to the
following prescription. We first generate $p$ patterns at random from a
Gaussian distribution with zero mean and unit variance. As in related
studies \cite{KraOp,DGP} we choose Gaussian patterns because they show less
pronounced finite size fluctuations. Next we use the Gray Code \cite{NR} to
run through all coupling vectors $\bf J$ and determine the corresponding
output strings. Finally we determine the size of the cells by counting the
multiplicity of the occurring outputs and compile a histogram of cell sizes in a double
logarithmic scale. The results are averaged over $10^4$ (for $N=10$) to $10$
(for $N=30$) realizations of the random patterns. Although the main
reason for the lesser number of realizations used for large $N$ 
was limited computer time it
became quite clear in the simulations that the sample-to-sample fluctuations
for large $N$ are due to self-averaging substantially smaller than for
small $N$. The results of the enumerations together with the corresponding
analytical results already displayed in fig.3 are shown in fig.4. It is 
at first surprising that the histograms lie always above the analytical
curves. However, for small $\gamma$ we have
\beq
\bay{lll}
2^{\gamma N} &= \int_0^1 d\alpha 2^{N f(\alpha)}\\[.2cm]
    &\simeq 2^{N f(\alpha_0)}\int_0^1 d\alpha
        \exp(\frac{N}{2}\;\log 2\;f''(\alpha_0)(\alpha-\alpha_0)^2)\\[.2cm]
       &\simeq  2^{N f(\alpha_0)} \sqrt{\frac{2\pi}{N\log 2 |f''(\alpha_0)|}}
\eay
\eeq
giving rise to
\beq\label{fss}
f(\alpha_0) = \gamma +\frac{1}{2N}\log\frac{N\log 2 \;|f''(\alpha_0)|}{2\pi}
\eeq
Hence the maximum of the histograms for finite $N$ converges 
to the asymptotic value
$f(\alpha_0)=\gamma$ for $N\to\infty$ from above. In fact in the inset of
fig.4 we have compared the finite size scaling predicted by (\ref{fss}) with
enumerations results for $f(\alpha_0)$ at $\gamma=.5$. The agreement is very good. 

\begin{figure}[htb]
 \epsfysize=8.5cm
       \epsffile{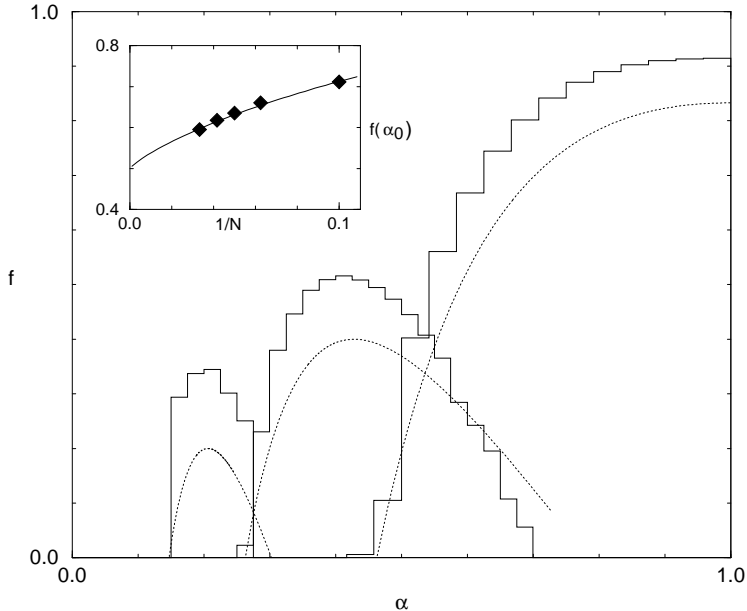}
\caption
{{Exact enumeration results for the multifractal spectrum $f(\alpha)$ of the}
Ising perceptron for
$p=6, N=30; p=12, N=30$ and $p=20, N=24$ (from left to right). The dotted
lines are the analytical results of fig.3 for $\gamma=0.2; 0.4$ and $0.833$
respectively. The inset shows a finite size scaling for $f(\alpha_0)$ at 
$\gamma=0.5$. The diamonds are enumeration results whereas the line is given
by eq.(\ref{fss}) with $|f''(\alpha_0)|$ estimated from the numerical data.
The rms fluctuations of the enumeration results are smaller than the symbol
size. The asymptotic value for $f(\alpha_0)$ is $0.5$.}
\end{figure}

\section{Summary}
In this paper, we have presented a multifractal analysis of the coupling
space of the single-layer perceptron with continuous and Ising couplings.
This has been done by characterizing the random partition of the coupling space into
different cells corresponding to different output sequences on the same fixed set
of random input vectors. This picture allowed us to refine the standard
Gardner analysis and, moreover, to unify the different approaches to
the storage problem of Gardner \cite{gardner} and Cover \cite{cover} and  
the generalization problem within one consistent picture. The different 
questions are related to different fractal subsets of the coupling
space: The cells with the most frequent size describe the storage problem, 
those dominating the total volume are related to the generalization ability for a
randomly drawn teacher. 

We have shown that the storage and the generalization problem can always 
be analyzed within the region where replica symmetry is locally stable. 
Moreover we have demonstrated 
that the most important part of the multifractal spectrum can be determined
within an annealed calculation with respect to the input pattern
distribution. This is not only an important technical advantage but may also
smooth the way for mathematically rigorous investigations of these problems.

Replica symmetry must be broken if one aims at describing comparatively
large and therefore rare cells. Despite the symmetry of the coupling space under point
reflection at the origin these cells no longer percolate in the sense that
it is impossible to reach the corresponding ``mirror'' cell without entering
cells of smaller size. This clustering of large cells is described by the
higher order parameters of a solution with broken replica symmetry.

Finally we note that the central procedure in our calculations is the
determination of {\it positive integer} moments $V^q$ of the distribution of
phase space volumes and the continuation of the result to real $q$ \cite{PaVi}.
In doing so we encountered divergences for all $q<0$. These are probably due to the
existence of empty cells $V=0$. We therefore hope that an appropriately
modified formalism describing the meta-stable state that occurs for $q<0$ might 
be able to yield also results for the VC-dimension of neural networks.

\vspace{1cm}
{\bf Acknowledgement:} Stimulating discussions with R\'emi Monasson and
Tam\'as T\'el are gratefully acknowledged.


\begin{thebibliography}{[99]}
\bibitem{rem2} The simplest case is given by a pattern set in which one
        pattern occurs at least twice since it is, of course, impossible to
        map identical patterns to different outputs.
\end{thebibliography}
\end{document}